\documentclass[conference]{IEEEtran}
\IEEEoverridecommandlockouts
\usepackage{cite}
\usepackage{amsmath,amssymb,amsfonts}
\usepackage{algorithmic}
\usepackage{listings}
\usepackage{graphicx,rotating}
\usepackage{textcomp}
\usepackage{xcolor}
\usepackage{hyperref}
\usepackage{subfigure}

\usepackage{braket}
\usepackage{tikz}
\usepackage{quantikz}
\usepackage{adjustbox}
\usepackage{multicol}

\usepackage{etoolbox}
\makeatletter
\patchcmd{\@makecaption}
  {\scshape}
  {}
  {}
  {}
\makeatother

\newif\iffinal
\finaltrue

\iffinal
 \newcommand{\raj}[1]{}
 \newcommand{\joaquin}[1]{}
 \newcommand{\TODO}[1]{}
 \else
 \newcommand{\raj}[1]{{\textcolor{blue}{ Raj: #1 }}}
 \newcommand{\joaquin}[1]{{\textcolor{brown}{ Joaquin: #1 }}}
 \newcommand{\TODO}[1]{{\textcolor{red}{ TODO: #1 }}}
 \fi

\def\BibTeX{{\rm B\kern-.05em{\sc i\kern-.025em b}\kern-.08em
    T\kern-.1667em\lower.7ex\hbox{E}\kern-.125emX}}
\begin{document}

\title{Towards Distributed Quantum Computing by Qubit and Gate Graph Partitioning Techniques
}

\definecolor{eclipseStrings}{RGB}{42,0.0,255}
\definecolor{eclipseKeywords}{RGB}{127,0,85}
\colorlet{numb}{magenta!60!black}

\lstdefinelanguage{json}{
    basicstyle=\normalfont\ttfamily,
    commentstyle=\color{eclipseStrings}, 
    stringstyle=\color{eclipseKeywords}, 
    numbers=left,
    numberstyle=\scriptsize,
    stepnumber=1,
    numbersep=8pt,
    showstringspaces=false,
    breaklines=true,
    frame=single,
    backgroundcolor=\color{gray}, 
    string=[s]{"}{"},
    comment=[l]{:\ "},
    morecomment=[l]{:"},
    literate=
        *{0}{{{\color{numb}0}}}{1}
         {1}{{{\color{numb}1}}}{1}
         {2}{{{\color{numb}2}}}{1}
         {3}{{{\color{numb}3}}}{1}
         {4}{{{\color{numb}4}}}{1}
         {5}{{{\color{numb}5}}}{1}
         {6}{{{\color{numb}6}}}{1}
         {7}{{{\color{numb}7}}}{1}
         {8}{{{\color{numb}8}}}{1}
         {9}{{{\color{numb}9}}}{1}
}

\author{\IEEEauthorblockN{Marc G. Davis$^{1}$, Joaquin Chung$^{2}$, Dirk Englund$^{1,3}$, Rajkumar Kettimuthu$^{2}$}
\IEEEauthorblockA{\textit{$^{1}$Massachusetts Institute of Technology, $^{2}$Argonne National Laboratory, $^{3}$Brookhaven National Laboratory} \\
mgd@mit.edu, chungmiranda@anl.gov, englund@mit.edu, kettimut@anl.gov}
}


\maketitle

\begin{abstract}
Distributed quantum computing is motivated by the difficulty in building large-scale, individual quantum computers.
To solve that problem, a large quantum circuit is partitioned and distributed to small quantum computers for execution.
Partitions running on different quantum computers share quantum information using entangled Bell pairs.
However, entanglement generation and purification introduces both a runtime and memory overhead on distributed quantum computing.
In this paper we study that trade-off by proposing two techniques for partitioning large quantum circuits and for distribution to small quantum computers.
Our techniques map a quantum circuit to a graph representation.  We study two approaches: one that considers only gate teleportation, and another that considers both gate and state teleportation to achieve the distributed execution.  Then we apply the METIS graph partitioning algorithm to obtain the partitions and the number of entanglement requests between them. 
We use the SeQUeNCe quantum communication simulator to measure the time required for generating all the entanglements required to execute the distributed circuit.
We find that the best partitioning technique will depend on the specific circuit of interest. 
\end{abstract}


\section{Introduction} \label{sec:intro}
Very recently, academics and companies building quantum computers have come to the realization that scaling to hundreds of thousands of qubits on a single chip is an extremely difficult task that will take longer than expected~\cite{quantum_interconnect}.
Thus, they are looking at distributed quantum computing as a promising technology to demonstrate useful quantum computing applications in the near future.
For instance, IBM has announced that their 462-qubit ``Flamingo'' processor (to be released in 2024), will leverage quantum communication between processors to support quantum parallelization~\cite{ibm-roadmap}.
Furthermore, in January of 2023, IonQ announced the acquisition of Entangled Networks stating that the acquisition will support their efforts to build large-scale quantum computers by enabling computation across multiple distributed quantum processors~\cite{ionq}.

Quantum computing problems are typically represented as quantum circuits that describe a series of quantum gates, measurements, and classical communications to be performed on a number of logical qubits.  
In the simplest case, these logical qubits are mapped one-to-one to physical qubits in a single quantum computer or quantum processing unit (QPU).
Throughout the paper we will use the terms QPU and quantum computer interchangeably.
Distributed quantum computing~\cite{dqc2004,rdv-dqc-2016,caleffi-dqc-2022} is the paradigm of using a set of small-capacity QPUs connected by a quantum network to solve a problem that otherwise would require a single, large-capacity quantum computer.
Networked quantum computers are composed of compute and communication qubits; the former execute the quantum circuit, while the latter establish Bell pairs between quantum computers during distributed execution.  Thus, quantum entanglement is the main enabler of distributed quantum computing.

Although distributed quantum computing can be seen as the salvation for large-scale quantum circuit execution, building an entanglement distribution network is not an easy task~\cite{ieqnet-arch2022, pompili2022experimental}.
Entanglement generation introduces overheads during distributed quantum computation execution~\cite{cnot_graph} in the form of qubits that must be involved in entanglement generation and purification, as well as time that must be spent performing these processes.  In particular, the qubit cost produces a range of quantum computer and circuit sizes for which distributed quantum computing might be beneficial, compared to single QPU computing.  If the reduction in number of qubits per QPU that are needed for computation is less than the number of qubits per QPU required for communication, then it will be better to simply run the circuit on a single QPU, as this will not incur the runtime cost.  Specifically, to observe a distributed quantum computing advantage in a given scenario involving a desired quantum circuit and a set of networked quantum computers, the quantum circuit must be big enough that none of the existing quantum computers could perform it independently, yet small enough that a sufficient amount of qubits are left for Bell pair generation (and purification) on each QPU once the circuit is divided. 
In this paper, we examine this trade-off and its implications for distributed quantum computing by assuming that each QPU in our quantum network has a limited number of qubits, some of which must be dedicated for communication.
Furthermore, we propose partitioning techniques and optimizations that may reduce the overhead of distributed quantum computing.
We use SeQUeNCe~\cite{sequence}, an event-drive simulator of quantum networks 
to determine if a given configuration would provide an advantage for distributed quantum computing.

\noindent The contributions of this paper are the following:
\begin{enumerate}
    \item We propose graph structures and partitioning techniques based on the mechanisms used to perform quantum gates on separate QPUs.
    \item We propose a graph post-processing procedure that ensures that partitions generated using our techniques respect the constraint that each QPU has a limited number of qubits.
    \item We present a preliminary evaluation of the overhead introduced by entanglement generation on distributed quantum computing by using SeQUeNCe. 
\end{enumerate}

The rest of the paper is organized as follows.
We provide background on the fundamental concepts of distributed quantum computing in Section~\ref{sec:background}.
We describe our graph partitioning and circuit mapping approaches in Section~\ref{sec:approach}, and we present our evaluation results in Section~\ref{sec:evaluation}.
We conclude in Section~\ref{sec:conclusion} with a brief summary and look at future work.

\section{Background} \label{sec:background}

Distributed quantum computing requires a process to map logical qubits onto the physical qubits in the quantum network, as well as techniques to perform interactions between logical qubits despite some qubits being represented on different QPUs.  Entanglement allows non-local operations between qubits in different quantum computers by means of quantum state or quantum gate teleportation.
The following subsections provide more details on the fundamental concepts of distributed quantum computing.
Section~\ref{sec:graph-part} describes how graph partitioning can be used to divide a large quantum circuit, while Sections \ref{sec:gen-entanglement} and \ref{sec:use-entanglement} describes how to generate and use entanglement in the context of distributed quantum computing, respectively.

\subsection{Graph Partitioning} \label{sec:graph-part}
Graph partitioning is the problem of partitioning the nodes of a graph into roughly equal partitions such that the total number (or weight) of edges split between partitions is minimized.
Many variants of this problem exist, but we will focus on the case where each pair of nodes has at most one edge, nodes are unweighted, and edges have integer weight.
The problem is then defined as follows: given a graph consisting of $n$ nodes and $v$ vertices, partition the nodes into $k$ partitions of approximately $\frac{n}{k}$ nodes each such that the sum of the weights of edges split between partitions is minimized.  
This problem is known to be NP-Hard, but high-quality approximate solvers for this type of problem have been developed~\cite{metis}.  
In this study we use the solver METIS~\cite{metis} to partition quantum circuits (see Section~\ref{sec:approach} for details).

\subsection{Generating Entanglement} \label{sec:gen-entanglement}
After a quantum circuit has been partitioned, distributed quantum computing requires the preparation of pairs of entangled qubits in the Bell state $\frac{1}{\sqrt{2}}(\ket{00}+\ket{11})$.  
These pairs of qubits are known as Bell pairs or e-bits.  
E-bits shared between nodes in the quantum network are a key resource that enables distributed quantum computing.  
The preparation of e-bits involves a technique to generate entanglement between distant physical qubits (e.g., entanglement heralding \cite{heralding}), usually over a photonic connection through fiber optic cables or free space.  
Producing e-bits over a long distance is a noisy process, and it results in low-quality entanglement.  
This can be improved by the use of entanglement purification~\cite{bennett1996purification}, which is a procedure that consumes many low-quality e-bits to produce a single higher fidelity e-bit.  
Entanglement purification incurs a cost in the form of run-time as well as requiring qubits to be dedicated for purification purposes~\cite{sequence}.

\subsection{Using Entanglement} \label{sec:use-entanglement}
Distributed quantum computing requires a way to perform quantum gates involving qubits on separate quantum computers.  
Fortunately, Bell pairs are a sufficient mechanism to perform universal distributed quantum computing.  
We will focus on two important techniques for utilizing Bell pairs to perform computations across multiple quantum computers:  
(1) quantum \textit{state} teleportation and (2) quantum \textit{gate} teleportation.  
Quantum gate teleportation enables distributed quantum computing through nonlocal $CNOT$ operations.

Which technique makes more efficient use of Bell pairs depends on the specifics of the quantum computing scenario.

\subsubsection{Quantum State Teleportation} \label{sec:state-teleport}
\begin{figure}
\begin{center}
\begin{adjustbox}{width=\columnwidth}
\begin{quantikz}
\lstick{$\ket{\psi}$} & \ctrl{1} & \gate{H} & \meter{} \vcw{+2} \\
\makeebit{$\frac{1}{\sqrt{2}}(\ket{00}+\ket{11})$} & \targ & \qw & \meter{} \vcw{+1} \\
& \qw & \gate{X} & \gate{Z} & \qw \rstick{$\ket{\psi}$}
\end{quantikz}
\end{adjustbox}
\end{center}
\caption{ 
This quantum circuit implements quantum state teleportation using a prepared Bell pair, or e-bit \cite{textbook}.
}
\label{fig:teleportcirc}
\end{figure}
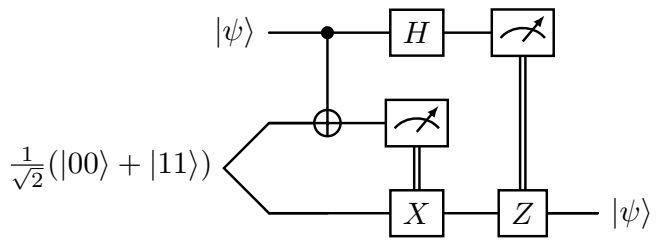

Quantum state teleportation is a technique that allows a logical qubit state to move between QPUs.  
This process works by interacting the qubit originally holding the state with one half of a Bell pair, followed by a series of measurements.
After these measurements are performed, the other half of the Bell pair will either be in a state equivalent to the original starting state, or one single-qubit operation away.  
The measurement results reveal which single-qubit operation must be applied to recover the original state.  
This process only requires Bell pair preparation and classical communication of measurement results (all other quantum operations are local).  
This method enables the transfer of a logical qubit state from one quantum computer to another, thus facilitating distributed quantum computing by sending qubits to the appropriate quantum computers for interactions.
A circuit for using quantum state teleportation to transport a single qubit state is shown in Figure \ref{fig:teleportcirc}.

\subsubsection{Quantum Gate Teleportation} \label{sec:gate-teleport}
Quantum teleportation does not require the entire state to be transferred to enable distributed quantum computing.  
The quantum teleportation process involves preparing an ancilla in a ``magic state'', interacting it with the original qubit, and measuring the original qubit.  
A 1-qubit operation will then be applied to the ancilla qubit, resulting in a state equivalent to performing a gate on the state from the original qubit.  
The gate performed is dependent on the specific magic state that was prepared.  
This technique is used in fault-tolerant quantum computing as a way to perform non-Clifford gates on error-corrected logical qubits.  
However, it is possible to teleport a $CNOT$ gate onto a pair of qubits.  
The magic state required is a Bell pair, which could be prepared between distant QPUs, effectively allowing a $CNOT$ to be performed between qubits on distant QPUs, without the need to send quantum states back and forth.

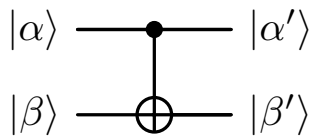
\begin{figure}
\begin{center}
\begin{adjustbox}{width=\columnwidth}
\begin{quantikz}
\lstick{$\ket{\alpha}$} & \ctrl{1} & \gate{H} & \gate{Z} \vcw{+1} & \qw \rstick{$\ket{\alpha '}$}\\
\makeebit{$\frac{1}{\sqrt{2}}(\ket{00}+\ket{11})$} & \targ & \qw & \qw & \meter{} \\
& \ctrl{1} & \gate{H} & \meter{} \vcw{+1} \\
\lstick{$\ket{\beta}$} & \targ & \qw & \qw & \gate{X} & \qw \rstick{$\ket{\beta '}$}\\
\end{quantikz}
\end{adjustbox}
\\
\texttt{\footnotesize The relationship between the states $\ket{\alpha}$, $\ket{\beta}$, $\ket{\alpha '}$, and $\ket{\beta '}$ is the same as would be produced from a normal CNOT gate:}
\\
\begin{adjustbox}{width=0.5 \columnwidth}
\begin{quantikz}
\\
\lstick{$\ket{\alpha}$} & \ctrl{1} \qw & \qw \rstick{$\ket{\alpha '}$}\\
\lstick{$\ket{\beta}$} & \targ \qw & \qw \rstick{$\ket{\beta '}$}
\end{quantikz}
\end{adjustbox}
\end{center}
\caption{ 
This quantum circuit implements quantum gate teleportation to perform a nonlocal CNOT using a prepared Bell pair, or e-bit \cite{Chou2018, Eisert2000}.
}
\label{fig:nonlocalgatecirc}
\end{figure}

A circuit for using quantum gate teleportation to implement a nonlocal CNOT is shown in Figure \ref{fig:nonlocalgatecirc}.

\section{Approach} \label{sec:approach}
Our approach to distributed quantum computing is to represent a large quantum circuit as a graph and then apply the METIS partitioning solver to determine the partitions that will be assigned to different QPUs.
The graph structure will depend on whether we are considering only quantum gate teleportation, or both quantum gate and state teleportation to realize the distributed quantum computation.
The following subsections provide details on our proposed graph structures and partitioning techniques (Sections \ref{sec:gate-graph} and \ref{sec:state-graph}), a graph post-processing procedure (Section~\ref{sec:graph-post}), and the quantum network simulator (Section~\ref{sec:sequence}) that we use to quantify the overhead introduced by the entanglement generation process.


\subsection{Qubit Partitioning} \label{sec:gate-graph}

When relying on nonlocal $CNOT$ gates as the method for achieving distributed quantum computing, the key metric for circuit optimization is to minimize the number of $CNOT$ gates between quantum computers.  This situation can be represented as a graph where the nodes are qubits and the edges are $CNOT$ gates, or equivalently, each edge is a pair of qubits between which at least one $CNOT$ is performed, and each edge has a weight equal to the number of $CNOT$s performed between that pair.  We want to optimize the quantum circuit by partitioning this graph such that the number of qubits in each partition is kept roughly equal, while the total edge weight between partitions is minimized.  The restriction that partitions remain equal in size allows the number of qubits needed per QPU to be minimized, while the goal of minimizing total edge weight equates to minimizing the number of $CNOT$ gates that must be performed nonlocally.
We call this technique \emph{Qubit Partitioning}, because each node represents a logical qubit.  Figure \ref{fig:qubit_partitioning} illustrates an example of Qubit Partitioning.



\subsection{Gate Partitioning} \label{sec:state-graph}

\begin{figure*}
    \centering
    \subfigure[Qubit Partioning: 4 e-bits used]{\label{fig:qubit_partitioning}\frame{\includegraphics[width=.49\textwidth]{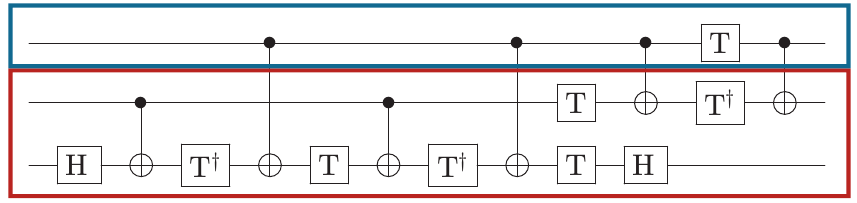}}}
    \hfill
    \subfigure[Gate Partitioning: 3 e-bits used]{\label{fig:gate_partitioning}\frame{\includegraphics[width=.49\textwidth]{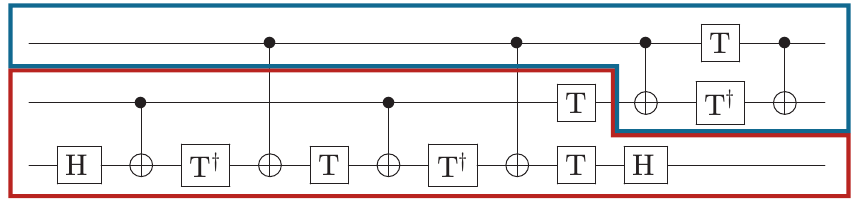}}}
	\caption{(a) shows an example of Qubit Partitioning of a quantum circuit. Two qubits are in the red partition and one qubit is in the blue partition. Four CNOT gates cross between partitions, and therefore must be performed nonlocally, meaning that four e-bits would be needed to perform this circuit in this way. (b) shows an example of Gate Partitioning of a quantum circuit.  Two qubits start in the red partition and one qubit starts in the blue partition, but partway through the circuit, one qubit is teleported from the red partition to the blue partition.  Two CNOT gates cross between partitions, and therefore must be performed nonlocally, and one wire crosses between two partitions, representing one qubit teleportation.  In total, three e-bits would be needed to perform this circuit in this way.}
	\vspace{-2ex}
	\label{fig:partitioning}
\end{figure*}

Nonlocal $CNOT$ gates are not always the most efficient mechanism for distributed quantum computing; sometimes state teleportation allows more efficient usage of Bell pairs.  To allow for the movement of logical qubits, the graph will be defined differently.  Each node will be a single qubit operation, or part of a two-qubit operation.  There will be two types of edges: those representing $CNOT$ gates, and those representing the chronological relation between two gates on the same qubit.  Partitioning a graph in this way has a different meaning.  The restriction that partitions have similar size represents the desire to have different QPUs on the network perform a similar amount of work.  The goal of minimizing the total weight of edges between partitions represents a desire to minimize the number of Bell pairs that must be used to complete the distributed circuit, regardless of whether those Bell pairs are used for gate or state teleportation.
We call this technique \emph{Gate Partitioning}, because each node represents a single-qubit gate or part of a $CNOT$ gate.  Figure \ref{fig:gate_partitioning} illustrates an example of Gate Partitioning.

\subsection{Graph Post-Processing} \label{sec:graph-post}

The graph structure that accounts for both gate and state teleportation is more general and therefore will produce more efficient results, in terms of Bell pair usage.  However, it has a flaw in that the condition of maintaining an equal amount of work on each quantum computer does not necessarily respect the requirement of each quantum computer having a limited number of available qubits.  This issue can be fixed with a post-processing routine, at the expense of adding additional state teleportations to the circuit.

Our post-processing procedure consists of traversing the circuit chronologically, and moving qubits to enforce a stronger condition on dividing qubits evenly.  This condition is expressed as an upper limit on the number of logical qubits that a quantum computer can have assigned at a time.  Whenever a qubit is moved via a quantum state teleportation, the procedure checks to see if this would violate the qubit limit on the receiving node.  If so, the procedure scans all qubits currently on that quantum computer, and finds the qubit move which will introduce the smallest number of additional nonlocal gates.  This process is repeated until the circuit can be completed without the qubit limit being violated by any state teleportation.

\subsection{SeQUeNCe} \label{sec:sequence}
For evaluating the effect of our graph partitioning schemes on realistic distributed quantum computing scenarios, we used the Simulator of QUantum Network Communication (SeQUeNCe)~\cite{sequence} to simulate how long it would take to produce a required number of e-bits under different assumptions of the available number of qubits, the fidelity of quantum entanglement heralding before any purification is done, and the target fidelity for purified e-bits.

SeQUeNCe is an open-source tool that allows modeling of quantum networks including photonic network components, control protocols, and applications.
The SeQUeNCe project website~\cite{sequence-github} provides the complete source code, documentation, and tutorials that allow easy usability and extendability. 
The simulator can be used to understand the trade-offs of alternative quantum network architectures, optimize quantum hardware, and develop a robust control plane. 
The modularized design of SeQUeNCe enables researchers to customize the simulated network and reuse existing models in a flexible way.
The Application module represents quantum network applications and their service requests. The resource management and network management modules use classical control messages to manage allocation of network resources. The entanglement management and hardware modules encompass protocols and hardware operations that act on qubits that are located in quantum memories or in photons at telecommunication wavelengths in optical network links. The SeQUeNCe simulator stores these states in the quantum state manager.
\section{Evaluation} \label{sec:evaluation}

To evaluate the proposed partitioning techniques, we ran circuit partitioning on a set of quantum circuits ranging from five qubits to 64 qubits.  We deliberately focus our evaluation on small circuits, because we foresee that our results could be verified on hardware in the near future.
We present our results in Table~\ref{tab:partition}.
We compare the \emph{Qubit Partitioning} technique and the \emph{Gate Partitioning} technique to the expected results from random qubit partitioning.  
We list the size of each circuit, in qubits and CNOTs, and then list the largest number of qubits required on an individual QPU and the number of e-bits required to implement the desired distributed circuit.  

\begin{table*}
\caption{
Summary of the results from running the two optimization strategies, compared to the expected result from simply randomly partitioning a graph using the Qubit Partitioning graph formalism.
}
\centering
\begin{tabular}{|c|c|c|c|c|}
\hline
Benchmark & Total Qubits & Total CNOTs & Max Qubits per Node & Required E-Bits\\
\hline
\hline
\multicolumn{5}{|c|}{Expected result from random qubit partitioning} \\
\hline
Grover & 5 & 48 & 3 & 30\\
TFIM-40 & 7 & 480 & 4 & 280\\
Adder & 9 & 98 & 5 & 56\\
QFT & 10 & 163 & 5 & 91\\
Multiplier & 10 & 216 & 5 & 120\\
Multiplier & 60 & 11405 & 30 & 5800\\
Adder & 63 & 1405 & 32 & 714\\
QFT & 64 & 5552 & 32 & 2821\\
\hline
\hline
\multicolumn{5}{|c|}{Qubit Partitioning (Nonlocal $CNOT$s only)} \\
\hline
Grover & 5 & 48 & 3 & 32\\
TFIM-40 & 7 & 480 & 4 & 240\\
Adder & 9 & 98 & 6 & 39\\
Qft & 10 & 163 & 7 & 72\\
Multiplier & 10 & 216 & 8 & 21\\
Multiplier & 60 & 11405 & 41 & 5064\\
Adder & 63 & 1405 & 33 & 715\\
QFT & 64 & 5552 & 37 & 2197\\
\hline
\hline
\multicolumn{5}{|c|}{Gate Partitioning (Nonlocal $CNOT$s and Teleportations)} \\
\hline
Grover & 5 & 48 & 3 & 32\\
TFIM-40 & 7 & 480 & 4 & 400\\
Adder & 9 & 98 & 5 & 57\\
QFT & 10 & 163 & 6 & 160\\
Multiplier & 10 & 216 & 6 & 135\\
Multiplier & 60 & 11405 & 31 & 5481\\
Adder & 63 & 1405 & 36 & 526\\
QFT & 64 & 5552 & 33 & 2409\\
\hline
\end{tabular}
\label{tab:partition} 
\end{table*}

After partitioning, we used {\tt SeQUeNCe} to benchmark the partitioned circuit to put the tradeoff of e-bit count vs. qubits used into context. 
For each circuit, we assumed that our quantum network consists of a pair of QPUs, each with 75\% of the number of qubits that would be needed to run the circuit on an individual quantum computer, rounded up.  For example, we assume quantum computers with four qubits when partitioning a five qubit circuit, and we assume 48 qubits-per-QPU when partitioning a 64 qubit circuit.  
We then simulated the time it would take to produce the requested number of Bell pairs of a specified quality with the allotted number of qubits.  
We used a target fidelity of 0.9, with two base fidelities: 0.85 (purification required) and 0.9 (no purification required).  We then ran the {\tt SeQUeNCe} simulator to estimate how long it would take to produce entanglement with a given base and target fidelity, and number of qubits available for purification.  We then multiplied that number by the number of e-bits required by the circuit, for an overall estimate of how long it would take to complete the circuit.  Here we assume that the time spent on quantum communication operations (i.e., entanglement generation and entanglement purification) is significantly larger than time spent on local quantum computing operations, thus the results obtained from {\tt SeQUeNCe} can be used as a estimate of the runtime achieved by running a quantum circuit in a distributed fashion.

\begin{figure}
\begin{center}
\includegraphics[width=\columnwidth]{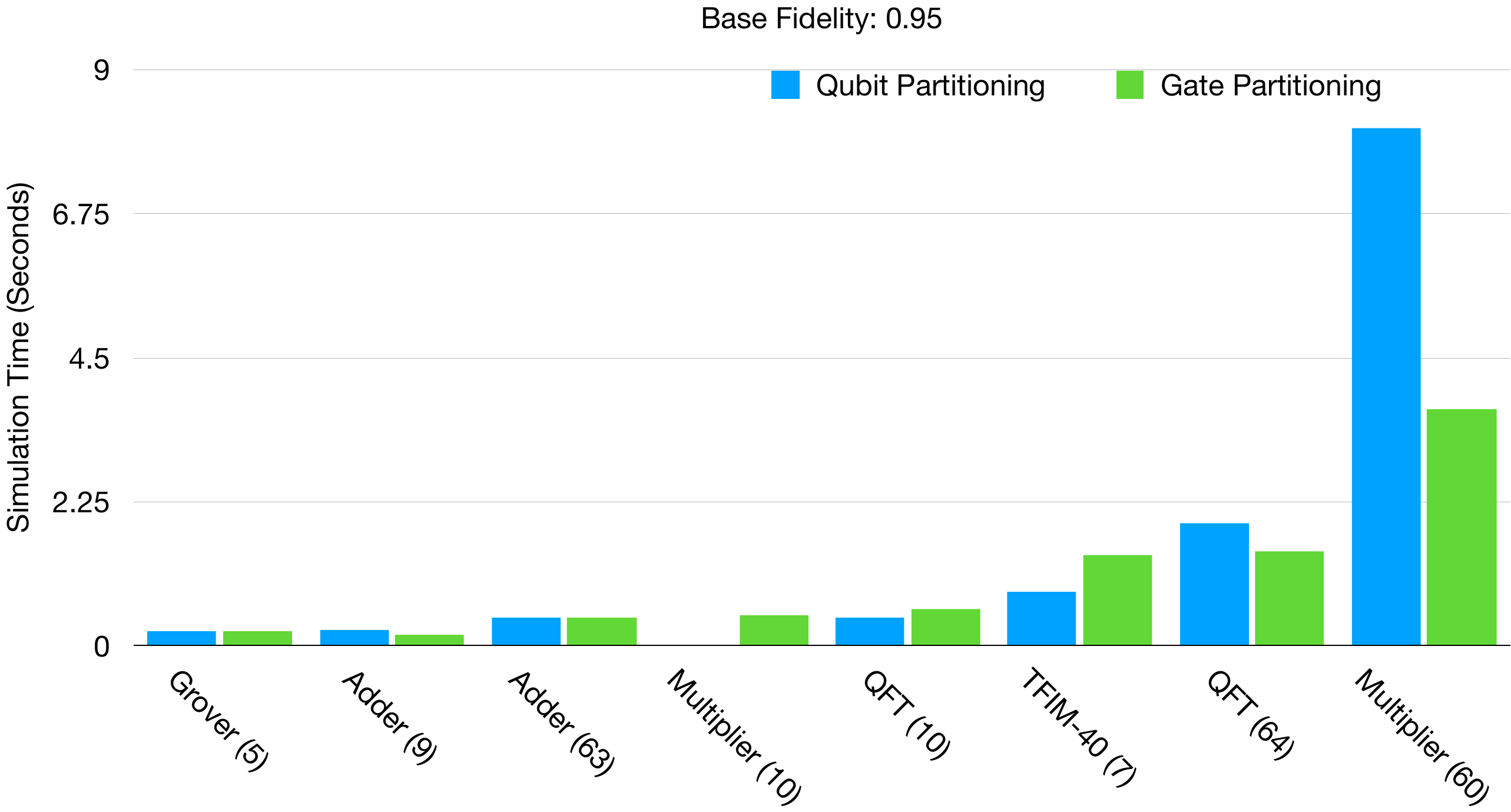}
\end{center}
\caption{ 
Comparison of simulated runtime of various circuits under both qubit partitioning (allowing only nonlocal CNOTs) and gate partitioning (allowing nonlocal CNOTs and qubit teleportations).  The post-processing technique described in Section \ref{sec:graph-post} was applied to the gate-partitioned circuits. The horizontal axis lists different quantum circuits, sorted in order of gate count.  Each circuit is named and has its size in qubits listed in parentheses.  The vertical axis displays the time (in seconds) that the simulation estimated it would take to execute the given partitioned circuit under the specifications described in Section \ref{sec:evaluation}.
}
\label{fig:bar_chart}
\end{figure}

\begin{figure}
\begin{center}
\includegraphics[width=0.8 \columnwidth]{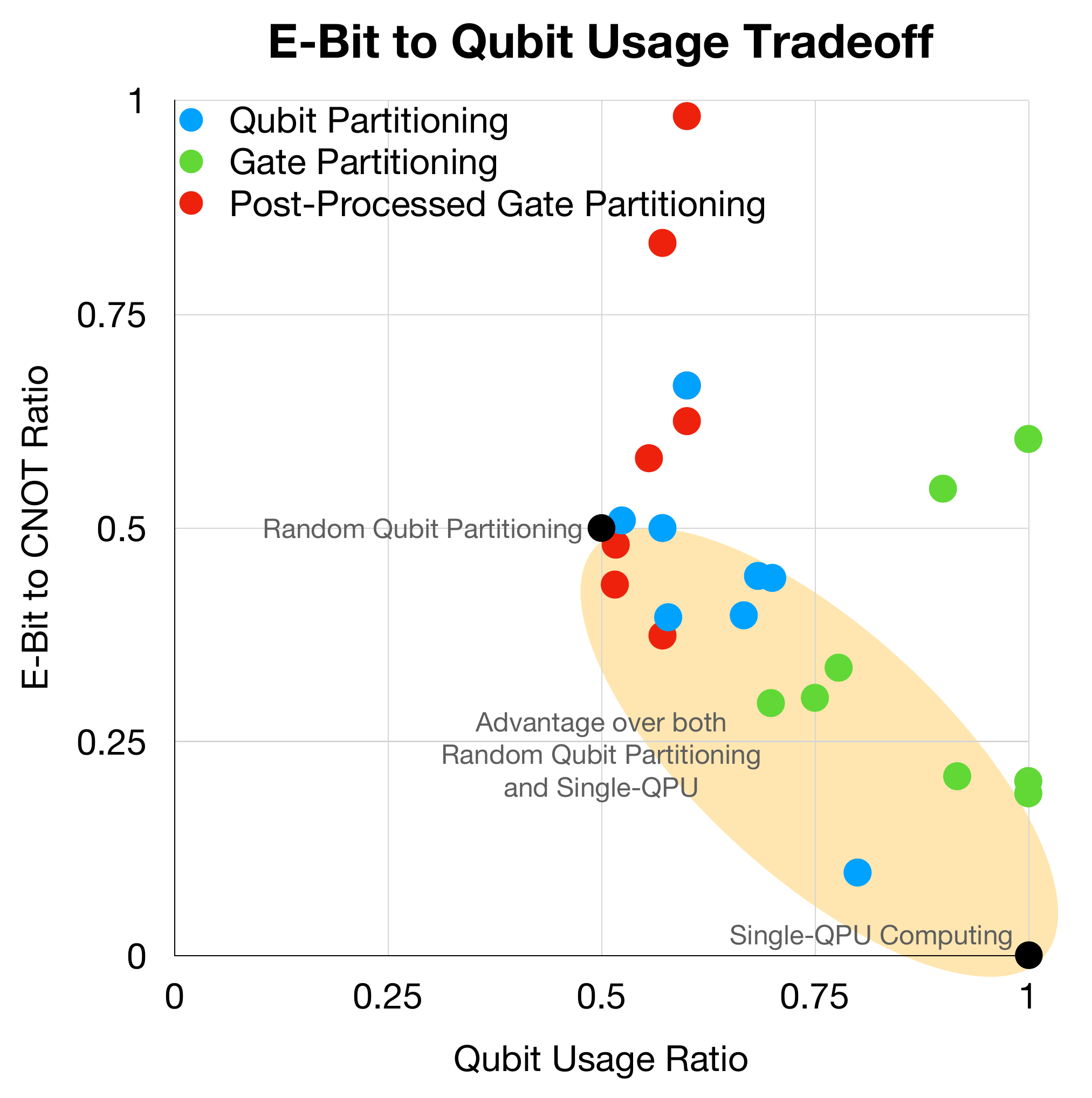}
\end{center}
\caption{
Comparison of the e-bit to qubit usage tradeoff between partitioning approaches.  The ``Qubit Partitioning'' approach allows only for nonlocal CNOTs, while the ``Gate Partitioning'' approach allows for both nonlocal CNOTs and qubit teleportations.  ``Post-Processed Gate Partitioning'' refers to gate-partitioning followed by the post-processing technique described in Section \ref{sec:graph-post}. ``Single-QPU Computing'' describes the situation where all of the qubits are in one partition, and no e-bits are used.  ``Random Qubit Partitioning`` involves randomly assigning exactly half of the qubits to one partition and half to the other, which should result in half of the CNOTs performed nonlocally on average.  To achieve an advantage over both random partitioning and single-QPU computing, a partitioned circuit must not have a largest partition of size equal to the number of qubits required by the original circuit, and must have an e-bit to CNOT ratio less than 50\%.  This region is highlighted in yellow.
}
\label{fig:pareto}
\end{figure}

Figure~\ref{fig:bar_chart} compares the simulated runtime of various circuits under both \emph{Qubit Partitioning} and \emph{Gate Partitioning} techniques.  Different procedures performed better in different situations.  Figure~\ref{fig:pareto} shows a Pareto-style comparison of all of the techniques we have studied, with random partitioning as a baseline.  In general, we would expect a random partitioning of a circuit to result in half of the gates being in each partition, and half of the CNOTs being performed nonlocally.  It is impossible to have fewer than half of the qubits in the largest partition, so the only way to achieve an advantage over the random-partitioning baseline is to reduce e-bit usage.  If one partition ends up being the size of the entire circuit, then there is no advantage to running the circuit in a distributed fashion.  These restrictions identify a region in which an advantage is obtained.  This region is highlighted in the yellow in Figure~\ref{fig:pareto}.  Not all partitioning approaches produced advantageous circuits, partially due to different techniques being better applicable to different circuits, depending on circuit structure.

 \section{Conclusion} \label{sec:conclusion}
Whether it is computing nodes in a quantum server room or distant quantum computers connected over a quantum internet, distributed quantum computing will play an important role in scaling quantum computation.  In this paper, we explored two different approaches to partitioning a quantum circuit for the purposes of distributed quantum computation, and we used a simulator of quantum networks to evaluate the overhead introduced by entanglement generation and purification.  We found that the best approach for circuit partitioning will depend on the specific circuit of interest.  In some cases, qubit partitioning, which allows only for nonlocal CNOTs, performs better, and in other cases gate partitioning, which allows for both nonlocal CNOTs and state teleportations, performs better.

\subsection{Future Work}

Our future work can be divided in three areas as follows: (1) verification of our simulation results experimentally, (2) increase in scale and complexity of our experiments, and (3) develop new algorithmic solutions to the problem of circuit mapping and graph partitioning.

\textbf{Verification of Simulation Results:} 
For this work, {\tt SeQUeNCe} was only used to estimate the overhead introduced by entanglement generation and purification procedures. To verify the results obtained with {\tt SeQUeNCe}, we will replicate the experiments in real quantum networking testbeds such as the one being built at the Chicago metropolitan area~\cite{ieqnet-arch2022}. However, verifying complete distributed quantum computing results is a more involved task as this will require an actual distributed quantum computing system that has not yet been built. Nevertheless, it could be possible to simulate two networked QPUs with a single large quantum computer. This can be done simply by designing a circuit that treats one set of qubits as one QPU and another set of qubits as another QPU, while only allowing interaction between these two ``virtual QPUs'' through a designated link. Moreover, one could possibly artificially add noise to such link by performing randomly chosen near-identity gates whenever the link is been used.

\textbf{Increase in Scale and Complexity of Experiments:}
Continuing our simulation work in the future, we plan to study more complex configurations such as considering the trade-offs of having a large number of small-capacity QPUs, or considering the effect of network topology, especially when a network is large enough to require quantum repeaters.
Another area of study is to scale the size and complexity of our input circuit. We intentionally focused on small circuits in this work as we expect that our results could be verified on hardware in short term. However, it would be interesting to study how these results translate to an impactful problem such as the Shor's algorithm when factoring a 2048-bit number.
We must reiterate that the graph partitioning problem is NP Hard, and the METIS solver only finds approximate solutions, thus we expect solution quality to decrease as the number of qubits increases. Nevertheless, processing large circuits within a reasonable runtime should still be feasible.
One strategy to tackle this problem could be to split up a larger algorithm (e.g., Shor's algorithm) into several smaller circuits such as adders or QFTs, like the ones studied in this work.  That approach may help mitigate the problem of scaling, although it may not completely solve it. This is an open area of research and approaches will evolve with the hardware that would be available in the future.

\textbf{New Algorithmic Solutions:}
The gate partitioning approach is more general than the qubit partitioning one, however it is harder to map gate partitioning onto a well-studied problem which is why it does not always provide an improvement in this work. Thus, for future work we would like to develop a more targeted algorithm that can take full advantage of the opportunities made available by gate partitioning.
For instance, one direction would be to model gate partitioning as a different type of problem, such as an integer linear programming problem (a more general type of problem than graph partitioning).  Finding a problem that can more closely represent our goals could lead to higher quality solutions without the need of a post-processing algorithm.

\section*{Acknowledgment}
The authors would like to thank Alex Kolar and Allen Zang for technical advice and support running the simulations in SeQUeNCe.
The authors would also like to thank the anonymous reviewers for their insightful questions and comments that helped improve this manuscript.  This material is based on work sponsored by the NSF, Contract Number PHY-1734011.  This material is based upon work supported by the U.S. Department of Energy, Office of Science, Office of Advanced Scientific Computing Research, Department of Energy Computational Science Graduate Fellowship under Award Number DE-SC0021110 and National Quantum Information Science Research Centers.

\section*{Disclaimer}
This report was prepared as an account of work sponsored by an agency of the United States Government. Neither the United States Government nor any agency thereof, nor any of their employees, makes any warranty, express or implied, or assumes any legal liability or responsibility for the accuracy, completeness, or usefulness of any information, apparatus, product, or process disclosed, or represents that its use would not infringe privately owned rights. Reference herein to any specific commercial product, process, or service by trade name, trademark, manufacturer, or otherwise does not necessarily constitute or imply its endorsement, recommendation, or favoring by the United States Government or any agency thereof. The views and opinions of authors expressed herein do not necessarily state or reflect those of the United States Government or any agency thereof.

\bibliographystyle{IEEEtran}
\bibliography{bibliography.bib}

\end{document}